\documentclass[fleqn,twoside]{article}%
\usepackage{amsmath}
\usepackage{amsfonts}
\usepackage{amssymb}
\usepackage{graphicx}
\usepackage{cite}
\def\be{\begin{equation}}
\def\ee{\end{equation}}
\def\bi{\bibitem}
\begin{document}
\title{Scalar meson field in a conformally flat space.}
\author{Abhik Kumar Sanyal$^1$ and D. Ray $^2$}
\maketitle
\noindent
\begin{center}
\noindent
$^1$ Dept of Physics, University College of Science,\\
92 A.P.C. Road, Calcutta-700009, India.\\
$^2$ Dept of Applied Mathematics, University College of Science,\\
92 A.P.C. Road, Calcutta-700009, India.\\
\end{center}
\footnotetext{\noindent
Electronic address:\\
\noindent
$^1$ sanyal\_ ak@yahoo.com \\
Present address: Dept. of Physics, Jangipur College, India - 742213.}
\noindent
\abstract{Among several authors, who studied massive and massless scalar meson fields in general relativity, attempts to obtain a complete set of solutions for a conformally flat metric $e^{\psi}\left({dx^1}^2 + {dx^2}^2 + {dx^3}^2 - {dx^4}^2\right)$ were made by Ray for massive and massless mesons and Gursay for massless mesons. Both of them concluded that $\psi$ must be a function of $K_0\left({dx^1}^2 + {dx^2}^2 + {dx^3}^2 -{dx^4}^2\right) + K_1 x^1 + K_2 x^2 + K_3 x^3 + K_4 x^4$, where,  where $K_0,~K_1,~ K_2,~K_3,~K_4$ are all constants. Both Ray and Gursay, however, overlooked an important particular case, which is studied here. As a by-product certain equations obtained by Auria and Regge in connection with ``Gravitational theories with asymptotic flat Instantons," are solved under less restrictive assumptions.}\\
\maketitle
\flushbottom
\section{Introduction:}

Einstein's field equations for general theory of relativity considering a scalar meson field to be the source, is given by
\be \label{1.1} R_{\mu\nu} - {1\over 2} g_{\mu\nu} R = - \left[\phi_{,\mu}\phi_{,\nu} - {1\over 2} g_{\mu\nu}(\phi_{,\alpha}\phi^{,\alpha} - m^2\phi^2)\right],\ee
in the unit $8\pi G = c = 1$, where $\phi$ is the meson field variable, $m$ is the associate mass, and $\phi_{,\mu} = {\partial \phi \over \partial x^\mu}$ and so on. The above equation \eqref{1.1} may also be expressed as,
\be\label{1.2} R_{\mu\nu} = - \phi_{,\mu}\phi_{,\nu} + {1\over 2} g_{\mu\nu} m^2 \phi^2.\ee
For a conformally flat metric,
\be\begin{split} & g_{\mu\nu} = e^\psi \eta_{\mu\nu}, ~~\mathrm{where}\\&
\eta_{\mu\nu} = 0, ~~\mathrm{for}~~\mu\ne \nu,\\&
\hspace{0.5 cm} = 1,~~\mathrm{for}~~\mu=\nu=1,2,3,\\&
\hspace{0.5 cm} = 1,~~\mathrm{for}~~\mu=\nu=4.\end{split}\ee
$R_{\mu\nu}$ takes the form,
\be\label{1.3} R_{\mu\nu} = \psi_{,\mu\nu} -{1\over 2} \psi_{,\mu}\psi_{,\nu} + {1\over 2}\eta_{\mu\nu}\chi,\ee
where,
\be\label{1.4} \chi = \psi_{,11} + \psi_{,22} +\psi_{,33} - \psi_{,44} + \psi_{,1}^2 + \psi_{,2}^2 + \psi_{,3}^2 - \psi_{,4}^2.\ee
The coupled equations \eqref{1.2} - \eqref{1.4} form the complete set of equations for a scalar meson field in a conformally flat space in general relativity. In order to simplify these equations and to study the properties of their solutions Ray \cite{1} noted that for $\phi =$ const the solution of the above-mentioned set of coupled equations is found quite simply, while for $\phi \ne$ const, the coupled equations lead to
\be\label{1.5a} \left(e^{-{\psi\over 2}}\right)_{,\mu} = \eta P_{,\mu} + \rho Q_{,\mu},\ee
\be\label{1.5b} {1\over 2}e^{-{\psi\over 2}} \phi_{,\mu} = \eta_{,\phi} P_{,\mu} + \rho_{,\phi} Q_{,\mu},\ee
\be\label{1.6a} 2\eta_{,\phi\phi} + \eta = \xi(\phi)\eta_{,\phi},\ee
\be\label{1.6b} 2\rho_{,\phi\phi} + \rho = \xi(\phi)\rho_{,\phi},\ee
where, $\eta = \eta(\phi)$ and $\rho = \rho(\phi)$, and \cite{2}
\be\label{1.7a} P = A x^1 + A_1 x^4,\ee
\be\label{1.7b} Q = {1\over 2}B\left({x^1}^2 + {x^2}^2 + {x^3}^2 - {x^4}^2\right) + B_1 x^1 + B_ x^2+ B_3 x^3 + B_4 x^4.\ee
In the above, $A, A_1, B, B_1, B_2, B_3, B_4$ are constants. From \eqref{1.5a} and \eqref{1.5b} Ray \cite{1} obtained
\be\label{1.8} 2 e^{-{\epsilon\over 2}}\left({\eta\rho_{,\phi}\over \eta_{,\phi}} - \rho\right) = {f(Q)\over e^{f(Q)\over 2}},\ee
where $\epsilon = \int{\eta d\phi \over \eta_{,\phi}}$, and $f$ is a function of $Q$. Since the left-hand side of Eq. \eqref{1.8} is a function of $\phi$ and the right-hand side of Eq. \eqref{1.8} is a function of $Q$, it was argued by Ray \cite{l} that from Eq. \eqref{1.8} one gets $\phi = \phi(Q)$. However, it is obvious from Eq. \eqref{1.8} that there arises another possibility which is
\be\label{1.9} 2 e^{-{\epsilon\over 2}}\left({\eta\rho_{,\phi}\over \eta_{,\phi}} - \rho\right) = {f(Q)\over e^{f(Q)\over 2}} = K_1,\ee
where $K_1$ is a constant. In this case $\phi$ need not be a function of $Q$. In the present paper we shall see whether this previously left out possibility of $\phi$ not being a function of $Q$ leads to
some new solutions. Also we explore whether the results obtained here can be suitably modified so as to be applied in the case of gravitational instantons (Sec. 3).

\section{New solutions for scalar meson field:}

In search of solutions that may have been left out by Ray \cite{1} we require that
\be \label{2.1} \eta_{,\phi} \ne 0, ~~\mathrm{and}~~{\rho_{,\phi}\over \eta_{,\phi}} \ne \mathrm{constant},\ee
and also
\be \label{2.2} Q_{,\mu} \ne 0,  ~~\mathrm{and}~~{P_{,\mu}\over Q_{,\mu}} \ne \mathrm{constant}.\ee
The reason for making these requirements is that, if Eq. \eqref{2.1} is not satisfied from Eq. \eqref{1.5b}, one gets
\be \phi = \phi(KP + K_1 Q),\ee
where $K$ and $K_1$ are constants. In view of the set of Eqs. (1.7), the above equation is essentially of the same form as $\phi = \phi(Q)$. Similarly if Eq. \eqref{2.2} holds, then  from Eq. \eqref{1.5b} one gets $\phi = \phi(Q)$. Since all the solutions of the form $\phi = \phi(Q)$ are already given by Ray \cite{1}, there is no need to look for such solutions. Now from the set of Eqs. (1.5) one can write,
\be \label{2.3} \begin{split} & \psi = \psi(P,Q), \hspace{0.7 in} \phi=\phi(P,Q),\\&
\left(e^{-{\phi\over 2}}\right)_{,P} = \eta,\hspace{0.6 in}\left(e^{-{\phi\over 2}}\right)_{,Q} =\rho,\\&
{1\over 2} e^{-{\phi\over 2}}\phi_{,P} = \eta_{,\phi},\hspace{0.5 in} {1\over 2} e^{-{\phi\over 2}}\phi_{,Q} = \rho_{,Q}.\end{split}\ee
Now, the $\phi =$ constant surfaces are given by,
\be\eta_{,\phi} dP + \rho_{,\phi} dQ = 0.\ee
Integrating, treating $\phi$ as a constant, we find
\be\label{2.4} \eta_{,\phi} P + \rho_{,\phi} Q = \lambda_{,\phi}.\ee
From Eqs. \eqref{2.1} and \eqref{2.2} we get
\be (\eta P + \rho Q)_{,\mu}= \lambda_{,\mu}\phi_{,\mu} + \left(e^{-{\psi\over 2}}\right)_{,\mu}.\ee
Integrating and absorbing the constant of integration we get
\be\label{2.5} \eta P + \rho Q = \lambda + e^{-{\psi\over 2}}.\ee
Differentiating Eq. \eqref{2.4} with respect to $P$, we get
\be (\eta_{,\phi\phi} P + \rho_{,\phi\phi}  - \lambda _{,\phi\phi})\phi_{,P} + \eta_{,\phi} = 0.\ee
From Eq. \eqref{2.3} we obtain
\be (2\lambda_{,\phi\phi}+\lambda) - (2\eta_{,\phi\phi}+\eta)P - (2\rho_{,\phi\phi} + \rho)Q = 0.\ee
From Eqs. \eqref{1.6a}, \eqref{1.6b} and \eqref{2.2} we get
\be \label{2.6} 2\lambda_{,\phi\phi} + \lambda = \xi(\phi) \lambda_{,\phi}.\ee
Clearly, from Eqs. \eqref{1.6a}, \eqref{1.6b} and \eqref{2.6}, $\eta,~\rho,~ p,~\mathrm{and}~ \lambda$ are found to be the solutions of the same ordinary linear second-order differential equation, and since in view of Eq. \eqref{2.1} $\eta,~\rho,~ \mathrm{and}~ P$ are linearly independent, there exists two constants $D$ and $D_1$  such that $\lambda = \eta D + \rho D_1$. Now from Eqs. \eqref{2.4} and \eqref{2.6}, we find
\be\label{2.7a} \eta_{,\phi}(P-D) + \rho_{,\phi}(Q-D_1) = 0,\ee
\be\label{2.7b} \eta(P-D) + \rho(Q-D_1) = e^{-{\psi\over 2}}.\ee
Further, Eq. \eqref{1.9} gives
\be 2 e^{-{\epsilon\over 2}}\left({\eta\rho_{,\phi}\over \eta_{,\phi}} - \rho\right) = K_1.\ee
Substituting $\eta$ and $\eta_{,\phi}$, from Eq. \eqref{2.7a} and \eqref{2.7b} and simplifying we get
\be \label{2.8} e^{\psi\over 2} = \left[{2\over K_1}(D_1 - Q)\right]e^{-{\epsilon\over 2}}.\ee
Now differentiating Eq. \eqref{2.7a} with respect to $\phi$ and using \eqref{1.5b}, we get
\be {1\over 2}e^{-{\psi\over 2}} + (P - D)\eta_{,\phi\phi} = (D_1 - Q)\rho_{,\phi\phi}.\ee
Now substituting $e^{-{\psi\over 2}}$ from Eq. \eqref{2.8} we have
\be \label{2.9}(P - D)\eta_{,\phi\phi} = (D_1 - Q)\left[\rho_{,\phi\phi} - {K_1 e^{\epsilon\over 2}\over 4}\right].\ee
From Eqs. \eqref{2.9}, \eqref{2.7a} and \eqref{2.7b} we get
\be\label{2.10} \eta_{,\phi\phi}\rho_{,\phi} = \eta_{,\phi}\rho_{,\phi\phi} - \eta_{,\phi} K_1\left({e^{\epsilon\over 2}\over 4}\right).\ee
Let us now choose
\be\label{2.11a} P - D = P_1,~~\mathrm{and}~~Q - D_1 = Q_1,\ee
so that
\be\label{2.11b} \alpha = {P_1\over Q_1},\ee
where,
\be\label{2.11c} P = A x^1 + A_1 x^4 - D,\ee
and
\be\label{2.11d} Q_1 = {1\over 2} B\left({x^1}^2+ {x^2}^2+ {x^3}^2 -{x^4}^2\right) + B_1 x^1 + B_2 x^2 + B_3 x^3 + B_4 x^4 - D_1,\ee
also
\be\label{2.11e} \phi = \phi(\alpha).\ee
Now,
\be \phi_{,\mu} = {\phi_{,\alpha}P_{1,\mu}\over Q_1} - \left({P_1\over Q_1^2}\right)\phi_{\alpha} Q_{1,\mu}.\ee
Comparing the above equation with Eq. \eqref{1.5b} and noting that $P$ and $P_1$ differ by a constant and likewise $Q$ and $Q_1$ we get
\be\label{2.12a} 2 e^{\psi\over 2} \eta_{,\phi} = {\phi_{,\alpha}\over Q_1},\ee
\be\label{2.12b}2 e^{\psi\over 2} \rho_{,\phi} = \phi_{,\alpha}\left({P_1 \over Q_1^2}\right).\ee
The general scalar meson field equation is
\be\label{2.13} -2 e^{\psi\over 2}\left(e^{-{\psi\over 2}}\right)_{,\mu\nu} = -\phi_{,\mu}\phi_{,\nu} + {1\over 2} g_{\mu\nu} m^2 \phi^2.\ee
From Eq. \eqref{2.12a} we get
\be\label{2.14} \left(e^{-{\psi\over 2}}\right)_{,\mu\nu} = {Q_{1,\mu\nu}\over F} - {F_{,\alpha}\over F^2}(Q_{1,\mu}\alpha_{,\nu}+ Q_{1,\nu}\alpha_{,\mu} + Q_{1,}\alpha_{,\mu\nu}  ) + {Q_1\alpha_{,\mu}\alpha_{,\nu}\over F^2}\left(2{F_{,\alpha}^2\over F} - F_{,\alpha\alpha} \right),\ee
where,
\be\label{2.15} F = {\phi_{,\alpha}\over 2 \eta_{,\phi}} = Q_1 e^{\psi\over 2}.\ee

\noindent
\textbf{Case-I:}\\
Let $\mu \ne \nu$. Equation \eqref{2.13} therefore becomes
\be\label{2.16} \alpha_{,\mu} \alpha_{,\nu}\left[{Q_1\over F^2}\left(2{F_{,\alpha}^2\over F} - F_{,\alpha\alpha}\right) - \phi_{,\alpha}^2{Q_1\over 2F}\right]= {F_{,\alpha}\over F^2}(Q_{1,\mu}\alpha_{,\nu}+Q_{1,\nu}\alpha_{,\mu}+ Q_{1}\alpha_{,\mu\nu}).\ee
Again, from the set of Eqs. (2.11) we get
\be (\alpha Q_1)_{,\mu\nu} = P_{1,\mu\nu} = 0, ~~~~~for~~~~~ \mu\ne \nu, \ee
which gives
\be Q_{1,\mu}\alpha_{,\nu}+Q_{1,\nu}\alpha_{,\mu}+ Q_{1}\alpha_{,\mu\nu} =0, ~~~~~for~~~~~ \mu\ne nu. \ee
Finally, from Eq. \eqref{2.16} we therefore get
\be\label{2.17} {1\over F}\left(2{F_{,\alpha}^2\over F} - F_{,\alpha\alpha}\right) - {1\over 2} \phi_{,\alpha}^2 = 0.\ee

\noindent
\textbf{Case-II:}\\
Let $\mu = \nu$. Using equation \eqref{2.13} and \eqref{2.17} it can easily be shown that the field equation for $\mu = \nu$ turns out to be
\be\label{2.18} {F_{,\alpha}\over F}(A - B\alpha) = B + {F^2m^2\phi^2\over 4Q_1}.\ee
Now, since $A,~ B~\mathrm{and}~ m$ are constants, $F$ and $\phi$ are functions of $\alpha$, while $Q_1$ is not a function of $\alpha$. So Eq. \eqref{2.18} can hold only if $m = 0$. So Eqs. \eqref{2.18} and \eqref{2.17} give
\be \label{2.19a}\begin{split}& F = {C\over A - B\alpha},\\&
\phi = 2\sqrt 2 \ln{(A - B\alpha)},\end{split}\ee
where,
\be \label{2.19b} \alpha = {P_1\over Q_1} = {A x^1 + A_1 x^4 - D \over {1\over 2}B\left({x^1}^2+{x^2}^2+{x^3}^2-{x^4}^2\right)+ B_1 x^1+  B_2 x^2+  B_3 x^3+  B_4 x^4 - D_1}.\ee
In the above, $A,~B$ and $C$ are all constants.

\section{Application to the case of gravitational theories with asymptotic flat Instantons:}

In a recent paper in connection with``Gravitational theories with asymptotic fiat instantons" Auria and Regge \cite{3} have sought to obtain a conformally fiat solution of
\be\label{3.1} R_{\mu\nu} - {1\over 2}g_{\mu\nu}R = {3\over 8}\left(\partial_{\mu}\phi\partial_{\nu}\phi - {1\over2}g_{\mu\nu}\partial^{\alpha}\phi\partial_{\alpha}\phi\right) - {3\over 8}g_{\mu\nu} M(\phi),\ee
where,
\be {\phi_{;\mu}}^{;\mu} = {1\over 2} {\partial M\over \partial \phi},\ee
with an Euclidean metric in the form $(+,+,+,+)$. Auria and Regge \cite{3} assumed
\be\label{3.2} g_{\mu\nu} = e^{\psi}\left({dx^1}^2 + {dx^2}^2 + {dx^3}^2 + {dx^4}^2\right),\ee
where,
\be\label{3.3} \psi = \psi\left({x^1}^2 + {x^2}^2 + {x^3}^2 + {x^4}^2\right).\ee
However, proceeding as in the paper by Ray \cite{1} and the present paper, it is easy to see that for solutions of Eq. \eqref{3.1} of the form Eq. \eqref{3.2})unless $M(\phi)= 0$, $\psi$ can be expressed as a function of either $\left({x^1}^2 + {x^2}^2 + {x^3}^2 + {x^4}^2\right)$ or $\left(K_lx^1 + K_2 x^2 + K_3 x^3 + K_4 x^4\right)$ where $K_l,~ K_2,~ K_3,~ K_4$ etc. are constants. However, if $\psi$ is a function of $\left(K_lx^1 + K_2 x^2 + K_3 x^3 + K_4 x^4\right)$, it is obvious that the solution cannot be asymptotically flat. Hence $\psi$  must be a function of $\left({x^1}^2 + {x^2}^2 + {x^3}^2 + {x^4}^2\right)$. So Eq. \eqref{3.3} is not an assumption but can be obtained automatically from other conditions.

\section{Conclusion:}

In summary, we find that Ray's conclusion that for massive and massless scalar meson fields in general relativity with a conformally fiat metric $e^{\psi}\left({dx^1}^2 + {dx^2}^2 + {dx^3}^2 - {dx^4}^2\right)$ that one must have $\psi$ as a function of $K_0\left({dx^1}^2 + {dx^2}^2 + {dx^3}^2 -{dx^4}^2\right) + K_1 x^1 + K_2 x^2 + K_3 x^3 + K_4 x^4$, is true except that another solution given by Eq. (2.19) is also possible for massless mesons. It should be noted that the case of massless mesons was also studied by Gursay \cite{4} and he also overlooked the above possibility. However, the calculations presented by G\"{u}rsay \cite{4} are too sketchy for us to find where the oversight occurred. Furthermore we find that the above result has significant application in the study of the ``Gravitational theories with asymptotically fiat instantons" by Auria and Regge \cite{3}, i.e., Eq. (3.3) is not an assumption but it naturally follows from Eqs. (3.1) and (3.2).

\end{document}